\documentclass[showpacs,preprintnumbers,amsmath,amssymb]{revtex4}
\usepackage{graphicx}% Include figure files
\usepackage{dcolumn}% Align table columns on decimal point
\usepackage{bm}% bold math
\usepackage{epsf}

\newcommand{\beq}{\begin{equation}}
\newcommand{\eeq}{\end{equation}}
\newcommand\beqa{\begin{eqnarray}}
\newcommand\eeqa{\end{eqnarray}}
\newcommand\bea{\begin{array}}
\newcommand\eea{\end{array}}
\newcommand{\nn}{\nonumber}
\newcommand{\neqa}{\nonumber\end{eqnarray}}
\newcommand{\la}{\label}

\newcommand{\eq}[1]{eq.(\ref{#1})}
\newcommand{\eqs}[2]{eqs.(\ref{#1},\ref{#2})}
\newcommand{\Eq}[1]{Eq.(\ref{#1})}

\newcommand{\ur}[1]{(\ref{#1})}

\newcommand{\Tr}{{\rm Tr}}

\newcommand{\Det}{{\rm Det}}
\newcommand{\half}{\frac{1}{2}}
\renewcommand{\d}{\partial}
\renewcommand{\O}{{\cal O}}

\renewcommand{\>}{{\rangle}}
\newcommand{\D}{r_{12}}

\newcommand{\bv}{\overline{{\rm v}}}
\newcommand{\Mphi}{\Phi}
\renewcommand{\v}{{\rm v}}

\newcommand{\re}{\relax{\rm I\kern-.18em R}}

\newcommand{\bshd}{\overline{\mathrm{sh}}\,}
\newcommand{\bchd}{\overline{\mathrm{ch}}\,}
\newcommand{\shd}{\mathrm{sh}\,}
\newcommand{\chd}{\mathrm{ch}\,}
\newcommand{\nablaslash}{\nabla\hspace{-.65em}/\hspace{.3em}}

\def\su2{{SU(2)}}

\def\tr{{\rm tr}}

\def\s{{\rm s}}
\def\r{{\rm r}}

\begin{document}
\title{Fermionic determinant for dyons and instantons with nontrivial holonomy}
\begin{abstract}
We calculate exactly the functional determinant for fermions in fundamental representation
of $SU(2)$  in the background of periodic instanton with non-trivial value of
the Polyakov line at spatial infinity.
The determinant depends on the value of the holonomy $\v$, the temperature,
and the parameter $\D$, which at large values can be treated as separation between
the Bogomolny--Prasad--Sommerfeld monopoles
(or dyons) which constitute the periodic instanton.
We find a compact expression for small and large $\D$ and
compute the determinant numerically for arbitrary $\D$ and $\v$.
%At large
%separation between constituent dyons, the quantum measure factorizes
%into a product of individual dyon measures, times a definite
%interaction energy.
\end{abstract}
\author{Nikolay Gromov$^a$}
\email{nik_gromov@mail.ru}
\author{Sergey Slizovskiy$^a$}
\email{ssliz@list.ru}
\vskip 0.3true cm
\affiliation{$^a$ St. Petersburg NPI, Gatchina, 188 300, St. Petersburg, Russia}
           % PACS numbers from http://publish.aps.org/PACS
           % 10. THE PHYSICS OF ELEMENTARY PARTICLES AND FIELDS
           % 11.15.-q Gauge field theories
           % 11.10.Wx Finite-temperature field theory
           % 11.15.Tk Other non perturbative techniques
\maketitle

\section{Introduction}
It is well-known that QCD with light fermions has a chiral symmetry
restoration phase transition at the temperature $T\simeq 170$ MeV.
This phase transition has been studied on the
lattice and with the help of various
QCD-motivated models (see, for example \cite{Kogut,Fleming}, \cite{Berges}), however, not much is
known about its driving mechanism.

When the temperature is high the effective coupling constant becomes
small and the theory can be studied semiclassically. From the other
side, at zero temperature it is very likely that the instanton --
anti-instanton ensemble (the instanton liquid) is responsible for
the spontaneous chiral symmetry breaking
\cite{DP2a,DP2b,DP3,Dobzor}. Therefore, in order to get an insight
into the mechanism of chiral symmetry breaking and restoration at
non-zero temperatures, it is natural to consider an ensemble of the
(anti) self-dual fields at finite temperature. A generalization of
the usual Belavin--Polyakov--Schwartz--Tyupkin (BPST)
instantons~\cite{BPST} for arbitrary temperatures is the Kraan--van
Baal--Lee--Lu (KvBLL) caloron with non-trivial
holonomy~\cite{KvB,LL}. These configuration were extensively
studied on the lattice in ~\cite{Brower,IMMPSV,Gatt} for $SU(2)$ and
$SU(3)$ gauge groups. To construct an ensemble of these
configurations theoretically, one should first compute the quantum weight of the
single KvBLL caloron, or the probability with which this
configuration occurs in the partition function of the theory. In
Ref.~\cite{DGPS} this problem was solved for the pure Yang--Mills
theory. To extend that result to the theory with fermions one has to
calculate the fermionic determinant in the background of the KvBLL
caloron. This calculation is the aim of the present paper.

Speaking of finite temperature one implies that the Euclidean
space-time is compactified in the `time' direction whose inverse
circumference is the temperature $T$, with the usual periodic
boundary conditions for boson fields and anti--periodic conditions
for the fermion fields. In particular, it means that the gauge field
is periodic in time, and the theory is no longer invariant under
arbitrary gauge transformations, but only under gauge
transformations that are periodical in time. As the space topology
becomes nontrivial the number of gauge invariants increases. The new
invariant is the holonomy or the eigenvalues of the Polyakov line
that winds along the compact 'time' direction~\cite{Polyakov}
\beq L=
\left.{\rm P}\,\exp\left(\int_0^{1/T}\!dt\,A_4\right)\right|_{|\vec
x|\to\infty}.
\la{Pol1}
\eeq\\
This invariant together with the topological charge and the magnetic
charge can be used for the classification of the field
configurations \cite{GPY} its zero vacuum average is
one of the common criteria of confinement.

The general expression for the self-dual electrically neutral
configuration with topological charge $1$ and arbitrary holonomy was
constructed a few years ago by Kraan and van Baal \cite{KvB} and Lee
and Lu \cite{LL}; it has been named the KvBLL caloron. In the
limiting case, when the KvBLL caloron is characterized by trivial
holonomy (meaning that the Polyakov line (\ref{Pol1}) assumes values
belonging to the group center $Z(N)$ for the $SU(N)$ gauge group),
it is reduced to the periodic Harrington-Shepard \cite{HS} caloron
known before. In this limit the quantum weights were studied in
detail by Gross, Pisarski and Yaffe \cite{GPY}.

Besides the neutral self-dual configurations there are charged
configurations. The ``elementary" soliton with unit electric and
magnetic charges is the dyon, also called the
Bogomolnyi--Prasad--Sommerfeld (BPS) monopole~\cite{Bog,PS}. It is a
self-dual solution of the Yang--Mills equations of motion with
static ({\it i.e.} time-independent) action density, which carries
both the magnetic and electric fields of the Coulomb type at
infinity, decaying as $1/r^2$. In the $SU(2)$ gauge theory there are
in fact two types of self-dual dyons~\cite{LY} $M$ and $L$ with
(electric,  magnetic) charges $(+,+)$ and $(-,-)$, and two types of
anti-self-dual dyons $\overline{M}$ and $\overline{L}$ with charges
$(+,-)$ and $(-,+)$, respectively. Formally, $L$ and $M$ dyons are
related by a non-periodical gauge transformation.

%%%%%%%%%%%%%%%%%%%%%%%%%%%%%%%%
% FIGURE 1
%%%%%%%%%%%%%%%%%%%%%%%%%%%%%%%%
\begin{figure}[t]
\centerline{ \epsfxsize=0.6\textwidth
\epsfbox{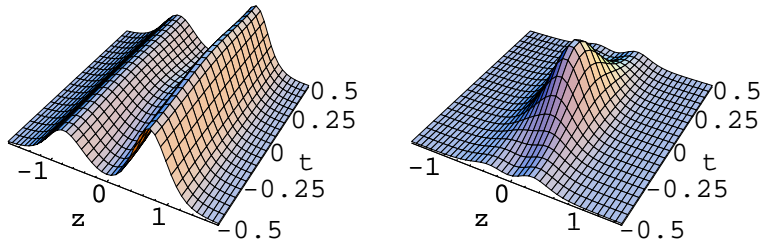}} \caption{The action density of
the KvBLL caloron as function of $z,t$ at fixed $x=y=0$, with the
asymptotic value of $A_4$ at spatial infinity $\v=0.9\pi T,\;
\bv=1.1\pi T$. It is periodic in $t$  direction. At large dyon
separation the density becomes static (left, $\D=1.5/T$). As the
separation decreases the action density becomes more like a $4d$
lump (right, $\D=0.6/T$). In both plots the L,M dyons are centered
at $z_{\rm L}=-\v\,\D/2\pi T,\;z_{\rm M}=\bv\,\D/2\pi T,\;x_{\rm
L,M}=y_{\rm L,M}=0$. The axes are in units of temperature
T.\label{fig:adp1}}
\end{figure}
%%%%%%%%%%%%%%%%%%%%%%%%%%%%%%%%

The KvBLL caloron of the $SU(2)$ gauge group (to which we restrict
ourselves in this paper) is ``made of" one $L$ and one $M$ dyon,
with total zero electric and magnetic charges. It means that for
large $\D$ the KvBLL caloron field becomes a sum of L and M dyons
fields ($\D$ is a parameter of the KvBLL caloron field, having a
natural meaning of the dyons' separation; it is associated with the
usual instanton size by $\D=\pi\rho^2 T$). One can consider the
KvBLL caloron as a two-monopole solution. Although the action
density of the isolated $L$ and $M$ dyons does not depend on time,
their combination in the KvBLL solution is generally non-static.
When the distance $\D$ becomes small compared to $1/T$ the KvBLL
caloron in its core domain $\sqrt{x_\mu x_\mu}\sim \rho$ reduces
to the usual BPST instanton (for explicit formulas
see \cite{KvB,DG}). The holonomy remains nontrivial as outside the
small core domain $A_4$ tends to a constant value.

The gluonic quantum weight computed in Ref.~\cite{DGPS} in the limit
when the separation between dyons $\D$ is much larger than their
core sizes $1/\v$ and $1/\bv$ (they are constrained by $\bv+\v=2\pi
T$) has the form: \beqa \nn {\cal Z}_{\rm KvBLL}&= &\int d^3z_1\,
d^3z_2\, T^6\, (2\pi)^{\frac{8}{3}}\,C\, \left(\frac{8 \pi^2}{g^2}
\right)^4 \left(\frac{\Lambda  e^{\gamma_E}}{4 \pi
T}\right)^{\frac{22}{3}} \left(\frac{\v}{2\pi T}\right)^{\frac{4
\v}{3\pi T}}
\left(\frac{\bv}{2\pi T}\right)^{\frac{4 \bv}{3\pi T}}\\
& \times &
\exp\left[-2\pi\,\D\,P''(\v)\right]\;\exp\left[-V^{(3)}P(\v)\right],
\eeqa where $P(\v)=\frac{\v^2\bv^2}{12\pi^2 T}$ and the overall
factor $C$ is a combination of universal constants; numerically
$C=1.0314$. $\Lambda$ is the scale parameter in the Pauli--Villars
regularization scheme; the factor $g^{-8}$ is not renormalized at
the one-loop level.

To account for fermions we have to multiply the partition function
by $\prod\limits_{j=1}^{N_f} \Det (i \nablaslash +i m_j)$, where
$\nablaslash$ is the spin-1/2 isospin-1/2 covariant derivative in
background considered, and $N_f$ is the number of light flavors. We
consider only the case of massless fermions here. The operator
$i\nablaslash$ has zero modes \cite{Cherndb} therefore a meaningful object is $\Det'(i
\nablaslash)$ --- a normalized and regularized product of non-zero
modes. In the self-dual background it is equal to $\left
(\Det(-\nabla^2) \right)^2$, where $\nabla$ is the spin-0
isospin-1/2 covariant derivative~\cite{BC}.

In this paper we calculate $\log\Det(-\nabla^2)$ exactly, find a
compact analytical expression for the large $\D$ asymptotic, find
$1/\D^n$ corrections up to the $5^{\rm th}$ order, and the small
$\D$ asymptotic. These results are analytical and give an almost
exact answer for all $\D$ except $\D\sim 1/T$. In section
\ref{sec_num} we present our results of numerical evaluation of the
determinant for arbitrary $\D$ and $\v$, which are consistent with
both asymptotics.

In the forthcoming publication~\cite{GSN} we are going to generalize our
results to arbitrary $SU(N)$. The first step in this direction has
been made in Ref.~\cite{DG} where the gluonic Jacobian over
zero-modes, or the volume form on the moduli space, has been
computed for the general $SU(N)$ KvBLL caloron (the metrics of the
caloron moduli space was found earlier by Thomas Kraan \cite{Kraan}). We believe that our
results will be useful to construct the ensemble of KvBLL calorons
at any temperatures and to obtain a better understanding of the
chiral symmetry restoration mechanism at the phase transition
temperature.

\section{The KvBLL caloron solution}

%%%%%%%%%%%%%%%%%%%%%%%%%%%%%%%%
% FIGURE 2
%%%%%%%%%%%%%%%%%%%%%%%%%%%%%%%%
%\begin{figure}[t]
%\centerline{
%\epsfxsize=0.4\textwidth
%\epsfbox{potential_energy.eps}}
%\caption{Potential energy as function of $\v/T$.
%Two minima correspond to $\half\Tr L=\pm 1$, the maximum corresponds to $\Tr L=0$.
%The range of the holonomy where dyons experience repulsion is shown in dashing.
%\label{fig:pot}}
%\end{figure}
%%%%%%%%%%%%%%%%%%%%%%%%%%%%%%%%
In this section we remind some basic facts about the caloron with
nontrivial holonomy just to establish the notations, see
Refs.~\cite{KvB,LL,DGPS} for a detailed discussion. We use here the
gauge convention and the formalism of Kraan and van Baal (KvB)
\cite{KvB}, the notations are taken from \cite{DGPS}.

The key quantity characterizing the KvBLL solution for the $SU(2)$
gauge group to which we restrict ourselves in this paper, is the
holonomy $\tr\,L$, \eq{Pol1}. In the gauge where $A_4$ is static and
diagonal at spatial infinity, i.e. $\left.A_4\right|_{{\vec
x}\to\infty}= i\v\frac{\tau_3}{2}$, it is this asymptotic value $\v$
which characterizes the caloron solution in the first place. We
shall also use the complementary quantity $\bv \equiv 2\pi T-\v$.
Their relation to parameters $\omega, \bar\omega$ introduced by
KvB~\cite{KvB} is $\omega= \frac{\v}{4\pi T},\; \bar\omega=
\frac{\bv}{4\pi  T}= \half-\omega$. Both $\v$ and $\bv$ vary from 0
to $2\pi T$. At $\v= 0,2\pi T$ the holonomy is said to be `trivial',
and the KvBLL caloron reduces to that of Harrington and
Shepard~\cite{HS}. Note that the fields in the fundamental
representation feel the sign of $\tr L$, therefore the cases $\v=0$
and $\v=2 \pi T$ differ when we account for fermions.

We shall parameterize the solution in terms of the coordinates of
the dyons' `centers' (we call the constituent dyons L and M
according to the classification in Ref.~\cite{DPSUSY}): \beqa \nn
L\;{\rm dyon}:\quad &&\vec z_1= -\frac{2\omega{\overrightarrow  \D}}{T},\\
\nn
M\;{\rm dyon}:\quad &&\vec z_2= \frac{2\bar\omega{\overrightarrow  \D}}{T},\\
\nn {\rm dyon\; separation}: &&\vec z_2-\vec z_1= {\overrightarrow
\D},\quad \quad |\D |= \pi T\,\rho^2, \eeqa where $\rho$ is the
parameter used by Kraan and van Baal; it becomes the size of the
caloron at $\v\to 0$. We introduce the distances from the
`observation point' $\vec x$ to the dyon centers, \beqa \nn
&&\vec r = \vec x-\vec z_1=\vec x + 2\omega\overrightarrow{\D},\qquad r= |\vec r|,\\
&&\vec s = \vec x-\vec z_2=\vec x -
2\bar\omega\overrightarrow{\D},\qquad s= |\vec s|. \eeqa Henceforth
we measure all dimensional quantities in units of temperature for
brevity and restore $T$ explicitly only in the final results.

The KvBLL caloron field in the fundamental representation
is~\cite{KvB} (we choose the separation between dyons to be in the
third spatial direction, ${\overrightarrow \D}= \D\vec e_3$): \beq
A^{\rm KvB}_\mu=\frac{i}{2}\bar\eta^3_{\mu\nu}\tau_3\d_\nu\log\Phi
+\frac{i}{2}\Phi\;{\rm
Re}\left[(\bar\eta^1_{\mu\nu}-i\bar\eta^2_{\mu\nu})(\tau_1+i\tau_2)\d_\nu\chi\right]
\label{APvB}\eeq where $\tau_i$ are Pauli matrices,
$\bar\eta^a_{\mu\nu}$ are 't~Hooft's symbols~\cite{tHooft} with
$\bar\eta^a_{ij}= \epsilon_{aij}$ and $\bar\eta^a_{4\nu}=
-\bar\eta^a_{\nu 4}= \delta_{a\nu}$. ``Re'' means hermitian part of
the matrix, and the functions used are \beqa \nn &&\chi=
\frac{\D}{\psi}\left(e^{-4\pi\bar\omega
ix_4}\frac{\shd}{s}+e^{4\pi\omega ix_4}\frac{\bshd}{r}\right),
\qquad\Mphi= \frac{\psi}{\hat\psi}\,
\label{eq:psihat},\\
&&\hat\psi= -\cos(2\pi x_4)+\bchd\chd+\frac{\vec r\vec
s}{2rs}\bshd\shd,\qquad \psi=
\hat\psi+\frac{\D^2}{rs}\bshd\shd+\frac{\D}{s}\shd\bchd+\frac{\D}{
r}\bshd\chd.\eeqa We have introduced the short-hand notations for
hyperbolic functions: \beq \shd\equiv\sinh(s\v),\qquad
\chd\equiv\cosh(s\v),\qquad \bshd\equiv\sinh(r\bv),\qquad
\bchd\equiv\cosh(r\bv) \,. \eeq One can note that the field
(\ref{APvB}) is not periodical. However, one can formally make a
(time dependent) gauge transformation so that the resulting field
\textit{is} periodical. It turns out that there are two inequivalent
possibilities to make a time dependent gauge transformation:
$g_P=e^{-2\pi i x_4\omega\tau_3}$ and $g_A=e^{2\pi i
x_4\bar\omega\tau_3}$. Correspondingly, we have two self-dual
periodical fields \beqa A^P_\mu(\omega,\vec x,x_4)=\delta_{\mu
4}2\pi i
\omega\tau_3+\frac{i}{2}\bar\eta^3_{\mu\nu}\tau_3\d_\nu\log\Phi
+\frac{i}{2}\Phi\;{\rm Re}\left[(\bar\eta^1_{\mu\nu}
-i\bar\eta^2_{\mu\nu})(\tau_1+i\tau_2)e^{-4\pi i x_4\omega}\d_\nu\chi\right]\label{APvBP}\\
A^A_\mu(\omega,\vec x,x_4)=-\delta_{\mu 4}2\pi i
\bar\omega\tau_3+\frac{i}{2}\bar\eta^3_{\mu\nu}\tau_3\d_\nu\log\Phi
+\frac{i}{2}\Phi\;{\rm
Re}\left[(\bar\eta^1_{\mu\nu}-i\bar\eta^2_{\mu\nu})(\tau_1+i\tau_2)e^{4\pi
i x_4\bar\omega}\d_\nu\chi\right] \label{APvBA}\eeqa Note that they
are related by an anti-periodical gauge transformation
$g_Pg_A^\dag=e^{-\pi i x_4\tau_3}$. In general, an anti-periodical
gauge transformation may affect different quantities. For example,
the determinant of the ghost operator in the fundamental
representation $\Det(-\nabla^2)$ is not invariant under this
transformation, whereas the determinant in the adjoint
representation  $\Det(-D^2)$ is invariant. We have to calculate
$\Det(-\nabla^2)$ for both fields (\ref{APvBP}) and (\ref{APvBA}).
Fortunately, in the case of one caloron we can limit ourselves to
one of the fields by the observation that these fields differ only
by the exchange of constituent dyons and by the substitution
$\omega\rightarrow\bar\omega$. More precisely, \beq
A^P_\mu(\omega,x_1,x_2,x_3,x_4)={\rm s_{\mu\nu}}\;
e^{-i\pi\frac{\tau_1}{2}}A^A_\nu(\bar\omega,x_1,-x_2,2[\omega-\bar\omega]\D-x_3,x_4)e^{i\pi\frac{\tau_1}{2}}
\eeq where $s$ is a diagonal matrix such that $\rm s_{11}=s_{44}=+1$
and $\rm s_{22}=s_{33}=-1$. We make a rotation around $\vec e_1$,
which exchanges the two dyons, and a global gauge transformation.
Obviously space rotation and gauge transformation of a background do
not change the determinant and we can conclude \beq
\Det\!\left(-\nabla^2[A^A_\mu(\omega)]\right)=\Det\!\left(-\nabla^2[A^P_\mu(\bar
\omega)]\right) \eeq This allows us to consider only $A^P_\mu$ in
what follows. We shall systematically drop the $P$ subscript.

The first term in \ur{APvBP} corresponds to a constant $A_4$
component at spatial infinity ($ A_4 \approx  i\v\frac{\tau_3}{2} $)
and gives rise to the non-trivial holonomy.

In the situation when the separation between dyons $\D$ is large
compared to both their core sizes $\frac{1}{\v}$ (M) and
$\frac{1}{\bv}$ (L), the caloron field can be approximated by the
sum of individual BPS dyons, see Figs.~1 (left). We give below the
field inside the cores and far away from both cores.

%%%%%%%%%%%%%%%%%%%%%%%%%%%%%%%%
% FIGURE 3
%%%%%%%%%%%%%%%%%%%%%%%%%%%%%%%%
%\begin{figure}[t]
%\centerline{
%\epsfxsize=0.6\textwidth
%\epsfbox{action_density_plot2.eps}}
%\caption{\label{fig:adp2} The action density of the KvBLL caloron as
%function of $z,x$ at fixed $t=y=0$. At large separations $r_{12}$ the caloron
%is a superposition of two BPS dyon solutions (left, $r_{12}=1.5/T$). At
%small separations they merge (right, $r_{12}=0.6/T$). The caloron
%parameters are the same as in Fig.~1.}
%\end{figure}
%%%%%%%%%%%%%%%%%%%%%%%%%%%%%%%%

\subsection{Inside dyon cores}

In the vicinity of the M dyon and far away from the L dyon ($r \bv
\gg1$) the field becomes that of the M dyon. It is static up to the
corrections of the order of $e^{-2\pi\bar\omega r}$ as can be seen
directly from \eq{APvBP} as $e^{-4\pi i x_4\omega}\chi$ become
static in this domain. We write it in spherical coordinates centered
at $\vec z_2$:
\[
A_4^M =\frac{i\tau_3}{2}\left(\v\coth(\v s)-\frac{1}{s}\right),\quad
A_\theta^M=\v\frac{\sin\!\phi\;\tau_1-\cos\!\phi\;\tau_2}{2i\sinh(\v
s)}\,,
\]
\beq A_r^M =0,\quad
A_\phi^M=\v\frac{\cos\!\phi\;\tau_1+\sin\!\phi\;\tau_2}{2i\sinh(\v
s)} +i\tau_3\frac{\tan(\theta/2)}{2s}, \la{Mdyon}\eeq
%whose asymptotics is
%\beqa
%\nn
%A^M_4 &\stackrel{r\to\infty}{\longrightarrow}& \left(\v-\frac{1}{s}\right)\,\frac{i\tau^3}{2},\;\;\;\;\;
%\nn A^M_\phi \stackrel{r\to\infty}{\longrightarrow} \frac{{\rm  tan}\frac{\theta}{2}}{s}\,
%\frac{i\tau^3}{2}
%\label{Mdyon_as}\eeqa

In the vicinity of the L dyon the field is substantially time
dependent. However, this time-dependence can be removed by an
\textit{anti-periodical} gauge transformation. It is instructive to
write the L dyon field in spherical coordinates centered at $\vec
z_1$: \beqa
\nn A_4^L&=&\frac{i\tau_3}{2}\left(\frac{1}{r}+2\pi-\bv\coth(\bv r)\right),\;\;\;A^L_r=0,\\
\nn A^L_\theta &=& i\bv\frac{-\sin(2\pi x_4-\phi)\;\tau_1
+\cos(2\pi x_4-\phi)\tau_2}{2\sinh(\bv r)},\\
A^L_\phi &=& i\bv\frac{\cos(2\pi x_4-\phi)\;\tau_1+\sin(2\pi
x_4-\phi)\tau_2}{2\sinh(\bv r)}-i\tau_3\frac{\tan(\theta/2)}{2r}\;.
\la{Ldyon}\eeqa We shall use the fact that the L dyon field can be
obtained from the M dyon field by interchanging $\v$ with $\bv$ and
by making an appropriate anti-periodical gauge transformation.

We see that in both cases the L,M fields become Abelian at large
distances, corresponding to the (electric, magnetic) charges $(-,-)$
and $(+,+)$, respectively. The corrections to the fields of M and L
dyons are hence of the order of $1/\D$, arising from the presence of
the other dyon.

\subsection{Far away from dyon cores}

Far away from both dyon cores ($r \bv\gg 1,\;s\v \gg 1$; note that
it does not necessarily imply large separations -- the dyons may
even be overlapping) one can neglect both types of exponentially
small terms, ${\cal O}\left(e^{-r\bv }\right)$ and ${\cal
O}\left(e^{-s\v}\right)$. With the exponential accuracy the function
$\chi$ in \eq{eq:psihat} is zero, and the KvBLL field \ur{APvB}
becomes Abelian~\cite{KvB}: \beqa \la{A4as}
&&A_4^{\rm  as}= \frac{i\tau_3}{2}\left(\v+\frac{1}{r}-\frac{1}{s}\right),\\
\la{Aphias} &&A_\varphi^{\rm  as}=
-\frac{i\tau_3}{2}\left(\frac{1}{r}+\frac{1}{s}\right)
\sqrt{\frac{(\D-r+s)(\D+r-s)}{(\D+r+s)(r+s-\D)}}\;. \eeqa In
particular, far away from both dyons, $A_4$ is the Coulomb field of
two opposite charges.

\section{The scheme for computing Det${\bf (-\nabla ^2)}$}

As explained in section I, to find the quantum weight of the KvBLL
caloron, one needs to calculate the small oscillation determinant,
${\rm Det}(-\nabla^2)$, where $\nabla_\mu= \d_\mu+A_\mu$ and $A_\mu$
is the caloron field \cite{KvB} in the fundamental representation.
We employ the same method as in \cite{DGPS,Zar}. Instead of
computing the determinant directly, we first evaluate its derivative
with respect to the holonomy $\v$, and then integrate the derivative
using the known determinant at $\v=0$ or $\v=2\pi$~\cite{GPY} as a
boundary condition.

If the background field $A_\mu$ depends on some parameter ${\cal
P}$, a general formula for the derivative of the determinant with
respect to such parameter is \beq \frac{\partial\,\log {\rm
Det}(-\nabla^2[A])}{\partial {\cal P}} = \!-\!\int
d^4x\,\Tr\left(\partial_{\cal P} A_\mu\, J_\mu\right),
\label{dvDet}\eeq where $J_\mu$ is the vacuum current in the
external background,  determined by the Green function: \beq
J^{ab}_\mu\!\equiv\!\left.(\delta^a_c\delta^b_d\d_x\!
-\!\delta^a_c\delta^b_d\d_y\!+\!A^{ac}\delta^b_d\!
+\!A^{db}\delta^a_c) {\cal G}^{cd}(x,y)\right|_{y= x}\qquad {\rm
or\;simply}\quad J_\mu\equiv \overrightarrow{\nabla}_\mu {\cal G}
+{\cal G} \overleftarrow{\nabla}_\mu. \label{defJ}\eeq Here $\cal G$
is the Green function or the propagator of spin-0,  isospin-1/2
particle in the given background $A_\mu$ defined by \beq
-\nabla^2_xG(x,y)= \delta^{(4)}(x-y) \la{Gdef}\eeq The anti-periodic
propagator can be easily obtained from it by a standard procedure:
\begin{equation}\label{greenP}
{\cal G}(x,y)= \sum_{n= -\infty}^{+\infty} (-1)^n G(x_4,{\vec
x};y_4+n,{\vec y}).
\end{equation}
\Eq{dvDet} can be easily verified by differentiating the identity
$\log{\rm Det}(-D^2)= \Tr\log(-D^2)$.  The background field $A_\mu$
in \eq{dvDet} is taken in the fundamental representation, as is the
trace. Hence, if the anti--periodic propagator ${\cal G}$ is known,
\eq{dvDet} becomes a powerful tool for computing quantum
determinants. Specifically, we take ${\cal P}= \v$ as the parameter
for differentiating the determinant, and there is no problem in
finding $\partial_{\v}A_\mu(r,s,\D,x_0,\v)$ for the caloron field
\ur{APvB}. We assume {\it positions of dyons fixed} for convenience,
it differs simply by a global translation from $\partial_{\v}A_\mu$
with the fixed center of mass position.

The Green functions in the self-dual backgrounds are generally
known~\cite{CWS,Nahm80} and are built in terms of the
Atiah--Drinfeld--Hitchin--Manin (ADHM) construction~\cite{ADHM} for
the given self-dual field. A subtlety appearing at nonzero
temperatures is that the Green function is defined by \eq{Gdef} in
the Euclidean $\re^4$ space, where the topological charge is
infinite because of the infinite number of repeated stripes in the
compactified time direction, whereas one actually needs an
explicitly anti--periodic propagator \ur{greenP}. To overcome this
nuisance, Nahm~\cite{Nahm80} suggested to pass on to the Fourier
transforms of the infinite-range subscripts in the ADHM
construction. We review this program in Appendix A of \cite{DGPS},
first for the single dyon field and then for the KvBLL caloron. This
way, we get the finite-dimensional ADHM construction both for the
dyon and the caloron, with very simple periodicity properties.
  The isospin-1/2 propagator in $\mathbb{R}^4$ is known to be
\beq
\label{green12} G(x,y)= \frac{\langle v(x)|v(y)\rangle}
{4\pi^2(x-y)^2}\;.
\eeq

In what follows it will be convenient to split it into two parts:
\beqa \nn
{\cal G}(x,y)&=& {\cal G}^\r(x,y)+{\cal G}^\s(x,y)\\
\la{green3} {\cal G}^\s(x,y)\equiv G(x,y),&&\qquad{\cal G}^\r(x,y)
\equiv\sum\limits_{n\neq 0} (\mp1)^n G(x_4,{\vec x};y_4+n,{\vec
y})\;. \eeqa The vacuum current \ur{defJ} will be also split into
two parts: ``singular'' and ``regular'' , in accordance to which
part of the periodic propagator \ur{green3} is used to calculate it:
\beq J_\mu= J^\r_\mu+J^\s_\mu. \eeq

%%%%%%%%%%%%%%%%%%%%%%%%%%%%%%%%
% FIGURE 4
%%%%%%%%%%%%%%%%%%%%%%%%%%%%%%%%
\begin{figure}[t]
\centerline{ \epsfxsize=0.5\textwidth \epsfbox{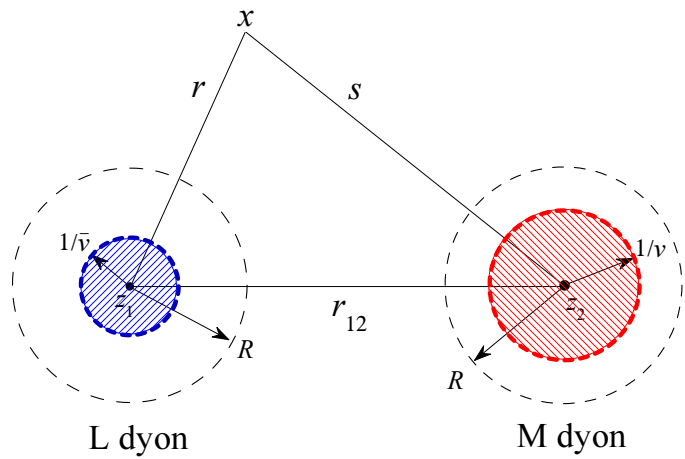}}
\caption{\label{fig:LMdyons} Three regions of integration for well
separated dyons.}
\end{figure}
%%%%%%%%%%%%%%%%%%%%%%%%%%%%%%%%

Although there is no principle difficulty in doing all calculations
exactly for the whole caloron moduli space, at the stage of spatial
integration of the determinant density we loose the capacity of
performing analytical calculations for the simple reason that the
expressions become too long, and so far we have not been able to put
them into a manageable form in the general case. Therefore, we have
to adopt a more subtle attitude. First of all we restrict ourselves
to the part of the moduli space corresponding to large separations
between dyons ($\D\gg 1$). Physically, it seems to be the most
interesting case. Furthermore, at the first stage we take
$\D\v,\D\bv\gg 1$, meaning that the dyons are well separated and do
not overlap since the separation is then much bigger than the core
sizes, see Fig.~1 (left). In this case, the vacuum current $J_\mu$
\ur{defJ} becomes that of single dyons inside the spheres of some
radius $R$ surrounding the dyon centers, such that
$\frac{1}{\v},\frac{1}{\bv}\ll R \ll \D$, and outside these spheres
it can be computed analytically with the exponential precision, in
accordance with subsection II.B, see Fig.~2. Adding up the
contributions of the regions near two dyons and of the far-away
region, we get $d\Det(-D^2)/d\v$ for well-separated dyons.
Integrating it over $\v$ we obtain the fermion determinant itself up
to a constant and possible $1/\D$ terms. Finally, we compute the
$1/\D$ corrections up to the $1/\D^5$ term in the $\Det(-D^2)$,
which turn out to be quite non-trivial.

This is already an interesting result by itself, however, we would
like to compute the constant, which can be done by matching our
calculation with that for the trivial caloron at $\v=0$ and $\v=2
\pi T$. It means that we have to extend the domain of applicability
to $\D\v=\O(1)$ (or $\D\bv=\O(1)$) implying overlapping dyons,
presented in Fig.~1 (right).
%It is simpler to make this extension, for the isospin-1/2 case for
%the reason that will become clear soon.

\section{Near each of the dyons } \la{nearcores}

In this section we calculate the right hand side of \eq{dvDet} in
the core domains of the L and M dyons. For M dyon this task is
nearly solved in \cite{DGPS}. In Appendix A the derivation of the
vacuum current in the background is given and we have to multiply it
by an expression for the gauge field (\ref{Mdyon}) and integrate
over M-dyon core domain (see Fig. 2).
The result is
 \beq
 \tr[\d_\v A^M_\mu J^M_\mu]=\frac{
       {\left(1-s\v \coth (s\v)\right) }^3
       \left( \sinh (2 s \v)-6 s \v \right)  }{96 {\pi }^2 s^3  \sinh^2(s\v)} -
   \frac{
     \left( 1-s\v \coth (s\v)\right)
     \left( \sinh (2 s\v)-2 s\v\right) }{24 s\sinh^2(s\v)}
\la{otvA} \eeq and integrating over the ball of radius $R$
 surrounding the dyon we obtain
\beq \int_0^R \tr[\d_\v
A^M_\mu J^M_\mu]d^4x=
\frac{\pi}{432}+\frac{\gamma_E}{12\pi}+\frac{31}{144\pi}+
\frac{\log(\v/\pi)}{12\pi} -\int^R \d_\v
P\left[\half\left(\bv+\frac{1}{s}\right)\right]4\pi s^2 ds\;,
\label{corA} \eeq where we denote
$P(\v)\equiv\frac{\v^2\bv^2}{12\pi^2}$ (in fact this
function is a one-loop perturbative effective potential \cite{GPY,NW})
and thus \beq -\int^R \d_\v
P\left[\half\left(\bv+\frac{1}{s}\right)\right]s^2 ds=
\frac{R^3}{6}P'\left(\frac{\bv}{2}\right)+\frac{R^2}{8}P''\left(\frac{\bv}{2}\right)
+\frac{R}{16}P'''\left(\frac{\bv}{2}\right)+\frac{\log
R}{96}P^{IV}\left(\frac{\bv}{2}\right) \eeq The calculation in the
L-dyon background is similar but slightly more difficult. The
strategy is to reduce this problem to the calculation in the M-dyon
background, which is simpler. For this purpose we note that the
L-dyon gauge field is a gauge transformation of the M-dyon gauge
field with the parameter $\v$ taken to be equal to $\bv=2\pi-\v$. We
should note, however, that this transformation is anti-periodical in
time and the anti-periodical Green function becomes a periodical one
under this gauge transformation. It means that we have to the use
the periodical vacuum current in the M-dyon background computed in
the Appendix A and substitute $\v$ with $\bv$ to get the right
answer: \beqa \tr[\d_\v A^L_\mu J^L_\mu]&=&
\frac{(r\bv\coth(r\bv)-1)(\sinh(2r\bv)-2r\bv)}{12 r\sinh^2(r\bv)}
+\frac{(r\bv\coth(r\bv)-1)^3(\sinh(2r\bv)-6r\bv)}{96\pi^2 r^3\sinh^2(r\bv)}\\
\nn&-&\frac{r\bv\coth(r\bv)-1}{\coth\left(\frac{r\bv}{2}\right)}\;
\frac{r\bv[2\cosh(r\bv)+\cosh(2r\bv)-4r\bv
\coth\left(\frac{r\bv}{2}\right)+7]-\sinh(2\r\bv)}{32\pi
r^2\sinh^2(r\bv)}. \eeqa Integration leads to the following result
\beq \int_0^R \tr[\d_\v A^L_\mu J^L_\mu]d^4x=
-\frac{\pi}{432}-\frac{\gamma_E}{12\pi}-\frac{31}{144\pi}-
\frac{\log(\bv/\pi)}{12\pi}-\int^R \d_\v
P\left[\half\left(\bv-\frac{1}{r}\right)\right]4\pi r^2 dr.
\label{corP} \eeq

Finally, substituting (\ref{corA}) and (\ref{corP}) into
(\ref{dvDet}) we arrive at the contribution from the cores \beq
\left.\d_\v\log\Det(-\nabla^2)\right|_{\rm
cores}=\frac{\log(\bv/\v)}{12\pi} -\frac{4\pi
R^3}{3}P'\left(\frac{\bv}{2}\right)-\frac{\pi
R}{2}P'''\left(\frac{\bv}{2}\right)\,.
% + \int^R \d_\v P\left[\half\left(\bv+\frac{1}{s}\right)\right]4\pi s^2 ds
%+ \int^R \d_{\v}  P\left[\half\left(\bv-\frac{1}{r}\right)\right]4\pi r^2 dr
\label{twocores} \eeq

\section{Contribution from the far region}\label{FarCurrent}

In this region we can drop the exponentially small terms $e^{-\v s}$
and $e^{-\bv r}$ in the r.h.s. of \eq{dvDet}. It turns out to be a
great simplification. The calculations are similarly to the calculations for
isospin-1 \cite{DGPS}. We are not repeating them here because the
result is simple and natural. The full vacuum current can be easily
expressed through the perturbative potential energy
$P(\v)\equiv\frac{\v^2\bv^2}{12\pi^2}$. The only non-zero component
is \beq
 J_4=-\frac{i \tau_3}{2} \left.P'(q)\right|_{q=\half\left ( \bv+\frac{1}{s}-\frac{1}{r}\right)}\,.
\la{J4asP}\eeq Making use of \eq{A4as} we come to the contribution
from the far region \beq \left.\d_\v\log\Det(-\nabla^2)\right|_{\rm
far}=\int \d_{\v} P\left[\half\left (
\bv+\frac{1}{s}-\frac{1}{r}\right)\right] d^4 x \eeq The integral is
taken over the whole space with two holes (see Fig.
\ref{fig:LMdyons}). We can use the symmetry between dyons to write
\beq \int \d_{\v} P\left[\half\left (
\bv+\frac{1}{s}-\frac{1}{r}\right)\right] d^4 x
=-\frac{1}{2}P'\left(\frac{\bv}{2}\right)\left(V-2\frac{4\pi}{3}R^3\right)
-\frac{1}{16}P'''\left(\frac{\bv}{2}\right)(4\pi \D-8\pi
R)+\O\left(\frac{\log R}{R}\right) \label{far1} \eeq where $V$ is
the 3D volume of the system.

\section{Combining all three regions}

Adding (\ref{far1}) to (\ref{twocores}) we get the expression for
the $\v$ derivative of the determinant for large distances $\D$
between constituents with the $1/\D$ precision: \beqa
\d_\v\log\Det(-\nabla^2)
=-\frac{\log(\v/\bv)}{12\pi}-\frac{1}{2}P'\left(\frac{\bv}{2}\right)V-
\frac{1}{16}P'''\left(\frac{\bv}{2}\right)4\pi \D \eeqa

This equation can be easily integrated over $\v$ up to a constant,
which in fact can be a function of the separation $\D$: \beq
\left.\log\Det(-\nabla^2)\right|_{\rm far}
=\left[P\left(\frac{\bv}{2}\right)-\frac{\pi^2}{12}\right]V+
P''\left(\frac{\bv}{2}\right)\frac{\pi\D}{2}-\frac{\bv\log{\bv}}{12\pi}-\frac{\v\log{\v}}{12\pi}+
f(\D) \label{result1} \eeq This result is valid with ${\cal
O}(\exp(-\v \D),\exp(-\bv \D),1/\D)$ precision, and we have
separated a constant $\frac{\pi^2}{12} = P(\pi),$  which accounts
for normalization to the $\v=0$ determinant. Since in the above
calculation of the determinant for well-separated dyons we have
neglected the Coulomb field of one dyon inside the core region of
the other, we expect that the unknown function
$f(\D)=\O(1/\D)+c_{\half}$ where $c_{\half}$ is the true integration
constant. One should consider the $\D$-derivative of the determinant
to see this explicitly. We perform it in Section \ref{sec1/r}. Our
next aim will be to find the constant $c_{\half}$. The $\O(1/\D)$
corrections will be found later.

\section{Expression for the constant in the determinant}

To find the integration constant, one needs to know the value of the
determinant at $\v\!=\!0$, where the KvBLL caloron with non-trivial
holonomy reduces to the Harrington--Shepard caloron with trivial
holonomy, and for which the determinant has been computed by Gross,
Pisarski and Yaffe (GPY)~\cite{GPY}. Before we match our determinant
at $\v\!\neq \!0$ with that at $\v\!= \!0$, let us recall the GPY
result for the isospin-1/2 determinant.

\subsection{Det${\bf (-\nabla^2)}$ at $\v=0$}

The $\v\!=\!0,\;2\pi$, periodic instanton is traditionally
parameterized by the instanton size $\rho$. It is
known~\cite{Rossi,GPY} that the periodic instanton can be viewed as
composed of two BPS monopoles one of which has an infinite size. It
becomes especially clear in the KvBLL construction~\cite{KvB,LL},
where the size of the M (L) dyon becomes infinite as $\v\to
0\;(2\pi)$, see section II. Despite one dyon being infinitely large,
one can still continue to parameterize the caloron by the distance
$\D$ between dyon centers, with $\rho= \sqrt{\D/\pi}$. Since the
determinant \ur{result1} is given in terms of $\D$ we have first of
all to rewrite the GPY determinant in terms of $\D$, too. Actually,
GPY have interpolated the determinant in the whole range of $\rho$
(hence $\D$) but we shall be interested only in the limit $\D\gg 1$.
In this range the GPY result reads:
\beq
\log\Det(-\nabla^2)=\left.\log\Det(-\nabla^2)\right|_{\v,T=0}+\eta
\frac{\pi\D}{3}-\frac{1}{12}\log\D+\frac{c_0}{16}+\O\left(\frac{1}{\D}\right)
\eeq
where $\eta=-1/2$ for anti-periodic boundary conditions (i.e.
$\v=0$, and M-dyon is infinitely large), $\eta=1$ for periodic
conditions (i.e. $\v=2\pi$, and L-dyon is large). One can verify
that the term linear in $d$ in this expression is exactly equal to
the term linear in $d$ in \eq{result1} when $\v=0$ for anti-periodic
and when $\v=2\pi$ for periodic boundary conditions. In
Ref.~\cite{DGPS} we have managed to find a constant $c_0$
analytically:
$$ c_0= \frac{8}{9}-\frac{8\,\gamma_E}{3}-\frac{2\,\pi^2}{27}+\frac{4\,\log \pi}{3} $$
The zero-temperature instanton determinant is~\cite{tHooft}:
\beqa
\la{Dinst} \left.\log \Det(-\nabla^2)\right|_{\v, T= 0}&= & \frac{1}{6}
\log \mu + \frac{1}{12} \log\left(\frac{\D}{\pi}\right) +\alpha(1/2)
\;,
\eeqa where it is implied that the determinant is regularized by
the Pauli--Villars method and $\mu$ is the Pauli--Villars mass. Thus
the isospin-1/2 result for the case of trivial holonomy is \beq
\log\Det(-\nabla^2)=\eta\frac{\pi\D}{3}+\frac{c_0}{16}+\alpha(1/2)-\frac{1}{12}\log\pi+\frac{1}{6}\log\mu\;,
\label{GPY2} \eeq where \beq
\alpha(1/2)=-\frac{17}{72}+\frac{\gamma_E}{6}+\frac{\log\pi}{6}-\frac{\zeta'(2)}{\pi^2}
\eeq
%Note, that no $\log\D$ is left.\\

We have to match our \eq{result1} with \eq{GPY2} at $\v=0$ or
$\v=2\pi$, but \eq{result1} has been derived assuming $\D\gg
\frac{1}{\v},\frac{1}{\bv}$ and one cannot formally take the limit
$\v\to 0$ in that expression without taking simultaneously
$\D\to\infty$. In the next subsection we will show that one
\textit{can} actually take this limit and get a correct constant. We
will rely on exact results to show that \eq{result1} is valid for
arbitrary values of $\v\D$ and $\bv\D$ (but large $\D\gg 1$).

\subsection{Extending the result to arbitrary values of $\v\D$}

Let us take a fixed but large value of the dyon separation $\D \gg
1$ such that both \eq{result1} and \eq{GPY2} are valid. Actually,
our aim is to integrate the exact expression for the derivative of
the determinant \beq \label{aim}
\partial_\v \log \Det(-\nabla^2)= \int \wp(x)d^4x ,\qquad
\wp(x)\equiv -\Tr\left[\partial_\v A_\mu J^\mu\right], \eeq from
$\v=0$, where the determinant is given by \eq{GPY2}, to some small
value of $\v\ll 1$ (but such that $\v\D\gg 1$), where \eq{result1}
becomes valid. We shall parameterize this $\v$ as $\v= k/\D\ll 1$
with $k \gg 1$. The result of the integration over $\v$ must be
equal to the difference between the right hand sides of
\eqs{result1}{GPY2}. We want to show that it is ${\cal O}(k/\D)$; it
means that there are no large corrections to \eq{result1} at small
$\v$ that can alter the constant term in the determinant.
% Of course, we must subtract the infra-red divergent terms that correspond to
% $P\left(\frac{\bv}{2}\right)V+
%P''\left(\frac{\bv}{2}\right)\frac{\pi\D}{2}$ in \eq{result1}
Denoting by $\bar{\wp}$ our $\wp$ with subtracted asymptotic terms
we have to show that \beq \int_0^{\frac{k}{\D}} d\v
\int\!d^4x\,\bar\wp(x)= \O\left(\frac{1}{\D}\right)\;.
\la{intwp2}\eeq
%The idea of the following is the same as in analysis of \cite{DGPS}.
In this integration we are in the domain $1/\v\gg 1$ and $\D\gg 1$
and we can simplify the integrand dropping terms which are small in
this domain. At this point it will be convenient to restore
temporarily the temperature dependence. With $\beta\equiv 1/T$ our
domain of interest is $1/\v\gg \beta$ and $\D\gg \beta$. Therefore
we are in the small-$\beta$ range and can expand $\bar\wp$ in series
with respect to $\beta$: \beq
\bar\wp=\frac{1}{\beta^2}\wp_0+\frac{1}{\beta}\wp_1+\O(\beta^0)\;.
\la{beta_exp}\eeq As we shall see in a moment, only the first two
terms can be not small in this range and we need to know only them
to compute $c_2$. It is a great simplification because $\wp_{0,1}$
does not contain terms proportional to $e^{-\bv r}$ since $\bv= 2\pi
T-\v \to \infty$ at $\beta \to 0$, and what is left is time
independent. Moreover, what is left after we neglect exponentially
small terms are homogeneous functions of $r,s,\D,\v$ and we can turn
to the dimensionless variables: \beqa &&\wp_0(r,s,\D,\v)=
\frac{1}{\D}\widetilde{\wp}_0\left(\frac{r}{\D},
\frac{s}{\D},\v\D\right) \;. \eeqa From the exact expressions we see
that $\wp_1 \equiv 0$. The reason is that there is no $1/\beta$ term
in \eq{Asum}. Actually the same phenomenon can be seen for the
single dyon in \eq{otvA}. We rewrite the l.h.s. of \eq{intwp2} in
terms of the new variables: \beqa
\int\!d^4x\,\bar\wp(x)=\frac{\D^2}{\beta}\int\! d^3\tilde
x\,\widetilde{\wp}_0 +\O(\beta) =\frac{\D^2}{\beta} F_0(\v \D) +
\O(\beta) \;, \la{intwp01}\eeqa where $\tilde x = x/\D$ is
dimensionless. We see that it is indeed sufficient to take just the
first two terms in the expansion \ur{beta_exp} at $\beta\to 0$.

Let us try to integrate \eq{intwp01} over $\v$ from 0 to $k/\D$ even
without explicit knowledge of $F_{0}$. We get \beq
  \int \limits_0^{k/\D}\frac{\D^2}{\beta} F_0(\v \D)
 d \v = \frac{\D}{\beta}  \int \limits_0^{k} F_0(\v\D) d(\v \D)
\eeq We know that the approximate result \eq{result1} is correct
with at least ${\cal O}(\exp(-k) , 1/(\v\D))$ precision, so at large
$k$ the correction to \eq{result1} must be a constant, plus a term
of the order of $1/(\v\D) = 1/k$. Therefore, $F_0$ must be zero. We
have checked numerically that it is indeed so (actually this
function $F_0$ is the same as in \cite{DGPS}).

Thus, we have proved \eq{intwp2}. We {\it can} now compare
\eq{result1} with \eq{GPY2} and get the constant: \beq
c_\half=\frac{c_0}{16}+\alpha(1/2)+\frac{1}{12}\log 4
\pi+\frac{1}{6}\log\mu=\frac{1}{6}\log\mu+
\frac{36\log(2\pi^2)-39-\pi^2}{216}-\frac{\zeta'(2)}{\pi^2}=\frac{1}{6}\log\mu+0.36584648
\label{c_half} \eeq

It means that \eq{result1} is right in whole range $0\leq \v\leq
2\pi$ with the accuracy $1/\D$.

\section{$1/r_{12}$ corrections} \label{sec1/r}

In the previous section we have seen that \Eq{result1} is valid to
all orders in $1/(\v\D),1/(\bv\D)$; however, there are other $1/\D$
corrections which are not accompanied by the $1/\v,\;1/\bv$ factors:
the aim of this subsection is to find them using the exact vacuum
current.

To this end, we again consider the case
$\D\gg\frac{1}{\v},\,\frac{1}{\bv}$ such that one can split the
integration over $3d$ space into three domains shown in Fig.~2. In
the far-away domain one can use the same vacuum current as we had in
Section \ref{FarCurrent}, as it has an exponential precision with
respect to the distances to both dyons. In the core regions,
however, it is now insufficient to neglect completely the field of
the other dyon, as we did in Section IV looking for the leading
order. Since we are now after the $1/\D$ corrections, we have to use
the exact field and the exact vacuum current of the caloron but we
can neglect the exponentially small terms of order $e^{-\D}$.

Another modification with respect to section IV is that we find it
more useful this time to choose $\D$ as the parameter $\cal P$ in
\eq{dvDet}. We shall compute the $1/\D^2$ and higher terms in
$\d\Det(\!-\!D^2)/\d\D$ and then restore the determinant itself
since the limit of $\D\to\infty$ is already known. Let us define how
the KvBLL field depends on $\D$. As is seen from \eq{APvBP} the
KvBLL field is a function of $r,\;s,\;\v,\;\D$ only. We define the
change in the separation $\D\to \D+d\D$ as the symmetric
displacement of each monopole center by $\pm d\D/2$. It corresponds
to \beq \frac{\d r}{\d \D}=\frac{\D^2+r^2-s^2}{4\D r},\qquad
\frac{\d s}{\d \D}=\frac{\D^2+s^2-r^2}{4\D s}. \la{drsd}\eeq We
shall use the definition \ur{drsd} to compute the derivative of the
caloron field \ur{APvB} with respect to $\D$.

Let us start from the $M$-monopole core region. To get the $1/\D$
correction to the determinant we need to compute its derivative in
the $1/\D^2$ order and expand correspondingly the caloron field and
the vacuum current to this order. Wherever the distance $r$ from the
far-away L dyon appears in the equations, we replace it by
$r=(\D^2+2s\D\cos\theta+s^2)^{1/2}$, where $s$ is the distance from
the M-dyon and $\theta$ is the polar angle seen from the M-dyon
center. Expanding in inverse powers of $\D$ we get the coefficients
that are functions of $s,\cos\theta$. One can easily integrate over
$\theta$ as the integration measure in spherical coordinates is
$2\pi s^2 ds\;d\cos\theta$. Then we integrate over $s$, the distance
to the core of the M-monopole. After that we have to add
contributions from the L and M dyon cores and from the far-away
domain. It turns out that contributions from dyons differ by terms
dependent on $R$ only: \beqa \d_{\D}\log\Det_{\rm core}&=&
\frac{1}{\D^2}\left(\frac{23}{72\pi}+\frac{\gamma}{6\pi}+\frac{\pi}{216}+\frac{\log(R^2
\v \bv/\pi^2)}{12\pi}\right)
+\frac{1}{12\D^3\pi}\left(\frac{1}{\v}+\frac{1}{\bv}\right)\\
\nn&-&\frac{1}{24\D^4\pi
}\left(\frac{1}{\v^2}+\frac{1}{\bv^2}\right)
+\frac{60-\pi^4}{2160\D^5\pi}\left(\frac{1}{\v^3}+\frac{1}{\bv^3}\right)+\O\left(\frac{1}{\D^6}\right)
\eeqa (we drop here powers of $R$ as they cancel with the
contribution from the far domain)

Now let us turn to the far-away domain. Recalling \eq{J4asP} we
realize that the contribution of this region is determined by the
potential energy: \beqa \left.\frac{\d \log
\Det(\!-\!\nabla^2)}{\d\D}\right|_{\rm far}
& = & \int\!d^3x\, \d_{\D}P\left(\half \left[\bv-\frac{1}{r}+\frac{1}{s}\right]  \right) \\
\nn& = & \frac{1}{8}P''(\bv/2)\int\!d^3x
\,\d_{\D}\left(\frac{1}{r}-\frac{1}{s}\right)^2 +\frac{1}{384}P^{\rm
IV}(\bv/2)\int\!d^3x
\,\d_{\D}\left(\frac{1}{r}-\frac{1}{s}\right)^4. \la{smpfn} \eeqa

The integration range is the $3d$ volume with two balls of radius
$R$ removed. We use \beqa \int
\d_{\D}\left(\frac{1}{r}-\frac{1}{s}\right)^2d^3x&=&4\pi-\frac{16\pi
R^2}{3 \D^2}
+\sum_{n=1}^\infty\frac{8 n \pi}{(2n-1)(2n+1)}\left(\frac{R}{\D}\right)^{2n+1}\\
\int \d_{\D}\left(\frac{1}{r}-\frac{1}{s}\right)^4d^3x&=&
\nn\frac{32\pi\log(\D/R)}{\D^2}+\frac{2\pi(8-9\pi^2)}{3\D^2}+\frac{80\pi
R}{\D^3} -\frac{32\pi R^2}{\D^4}+\frac{304\pi R^3}{15
\D^5}+\O(1/\D^{6})
%-\frac{24\pi R^4}{d^6}+\frac{2976\pi R^5}{175 d^7}\\
%&&-\frac{64\pi R^6}{3 d^8}+\frac{2720\pi R^7}{147 d^9}-\frac{20\pi R^8}{d^{10}}+\O(1/d^{11})
\eeqa
Adding up all three contributions we see that the region
separation radius $R$ gets canceled (as it should), and we come to:
\beqa
\d_{\D}\log\Det(-\nabla^2)&=&\frac{\pi}{2}P''\left(\frac{\bv}{2}\right)
+\frac{1}{12\D^2\pi}\left(\log(\v\bv\D^2/\pi^2)-\frac{23\pi^2}{72}+2\gamma+\frac{25}{6}\right)\\
&+&\frac{1}{12\D^3\pi}\left(\frac{1}{\v}+\frac{1}{\bv}\right)
-\frac{1}{24\D^4\pi }\left(\frac{1}{\v^2}+\frac{1}{\bv^2}\right)
\nn+\frac{60-\pi^4}{2160\D^5\pi}\left(\frac{1}{\v^3}+\frac{1}{\bv^3}\right)+\O\left(\frac{1}{\D^6}\right)
\;. \eeqa Integrating over $\D$ gives \beqa
\la{r12corr}\log\Det(-\nabla^2)&=&\frac{\pi}{2} P''\left
(\frac{\bv}{2}\right) \D -
\frac{1}{12\D\pi}\left(\log(\v\bv\D^2/\pi^2)-\frac{23\pi^2}{72}+2\gamma+\frac{37}{6}\right)
-\frac{1}{24\D^2\pi}\left(\frac{1}{\v}+\frac{1}{\bv}\right)\\
\nn&+&\frac{1}{72\D^3\pi
}\left(\frac{1}{\v^2}+\frac{1}{\bv^2}\right)
-\frac{60-\pi^4}{8640\D^4\pi}\left(\frac{1}{\v^3}+\frac{1}{\bv^3}\right)
+\frac{12-\pi^4}{2880\D^5\pi}\left(\frac{1}{\v^4}+\frac{1}{\bv^4}\right)
+\bar c+\O\left(\frac{1}{\D^6}\right)\;, \eeqa where $\bar c$ is the
integration constant that does not depend on $\D$. Comparing
\eq{r12corr} with \eq{result1} at $\D\to\infty$ we conclude that
\beq \bar c =
V\,\left[P\left(\frac{\bv}{2}\right)-\frac{\pi^2}{12}\right]
-\frac{\bv\log{\bv}}{12\pi}-\frac{\v\log{\v}}{12\pi}+c_\half \eeq
and $c_\half$ is given in \eq{c_half}. We note that the leading
correction, $\log\D/\D$, arises from the far-away range and is
related to the potential energy, similar to the leading $\D$ term.
This series in $1/\D$ is asymptotic as the coefficients
are rapidly increasing.

Interesting, we have revealed that the contribution from the regular
part of the current $J^{\rm r}$ has no corrections $1/\D^n$ (we
verified that up to the $10^{\rm th}$ order) and is determined
completely by the far asymptote \beq \!-\!\int
d^4x\,\Tr\left(\partial_{\D} A_\mu\, J^{\rm r}_\mu\right)\simeq
P''\left(\frac{\bv}{2}\right)\frac{\pi}{2}\;. \la{regcontr1} \eeq It
seems that $\Tr\left(\partial_{\D} A_\mu\, J^{\rm r}_\mu\right)$ is
a full derivative of a simple expression.

\section{Asymptote of small separation }

In this domain of the moduli space of the KvBLL caloron it is
important to realize that KvBLL caloron is a chain of the usual BPST
instantons equally separated in time direction with a fixed relative
gauge orientation between the neighbors.

When the size of instantons is large compared to their separation
$1/T$, they overlap and cannot be thought of as individual
pseudoparticles. It is much simpler to consider the KvBLL caloron as
a bound state of two BPS dyons. Small separations correspond to the
small instanton size $\rho$ since $\rho=\sqrt{\D/\pi}$. It means
that instantons do not overlap, and the KvBLL caloron reduces to the
usual BPST caloron.

Formally, one switches to dimensionless variables $r=\bar
r\rho,\;s=\bar s\rho$ and then takes the limit $\rho\rightarrow 0$.
In this limit the gauge field of the KvBLL caloron becomes that of
the ordinary instanton of vanishing size, \textit{plus} a constant
field. It is then clear that the determinant becomes that of the
ordinary instanton, plus a contribution from a constant field: \beq
\log \Det(-\nabla^2)=\left[P\left(\frac{\bv}{2}\right)
-\frac{\pi^2}{12}\right]V+\frac{1}{6} \log \mu + \frac{1}{12}
\log\left(\frac{\D}{\pi}\right) +\alpha(1/2)+\O(\rho)
\;,\;\;\;\;\;\alpha(1/2)=0.145873 \eeq One can expand in powers of
$\rho$ further, but it requires knowledge of the exact expression
for the vacuum current (\ref{defJ}). Here we give only the final
result. The contribution from the regular part of the current
$J^{\rm r}$ for small $\D$ is \beq \!-\!\int
d^4x\,\Tr\left(\partial_{\D} A_\mu\, J^{\rm r}_\mu\right)
=P''\left(\frac{\bv}{2}\right)\frac{\pi}{2}+\O(\rho)
\la{regcontr}\eeq and from the singular part
\beq
\!-\!\int d^4x\,\Tr\left(\partial_{\D} A_\mu\, J^{\rm s}_\mu\right)=\frac{1}{12\pi\rho^2}+(3\v\bv-2\pi^2)\frac{1}{72\pi}+\O(\rho)
\la{singcontr}
\eeq
 taking into account these contributions we have
\beq
\label{small_rho}
\log \Det(-\nabla^2)=\left[P\left(\frac{\bv}{2}\right)-\frac{\pi^2}{12}\right]V+
\alpha(1/2)+\frac{1}{6} \log(\mu\rho) -(7\pi^2-3\pi\v-3\v^2)\frac{\rho^2}{36}+\O(\rho^3) \;,
\eeq
Note that the contribution from the regular part of the vacuum current
for small values of $\D$ is exactly the same as it has been for
large $\D$ in \eq{regcontr1}. It suggests that \eq{regcontr1} is
exact. Unfortunately, we have not been able to prove it
analytically, but we have verified it numerically. We have found out
that \eq{regcontr1} is correct for all $\D$ and $\v$ with the
precision of at least $10^{-3}$.

\section{Numerical evaluation} \la{sec_num}

\begin{figure}[t]
\la{fig_2sht}
\centerline{\epsfxsize=0.5\textwidth \epsfbox{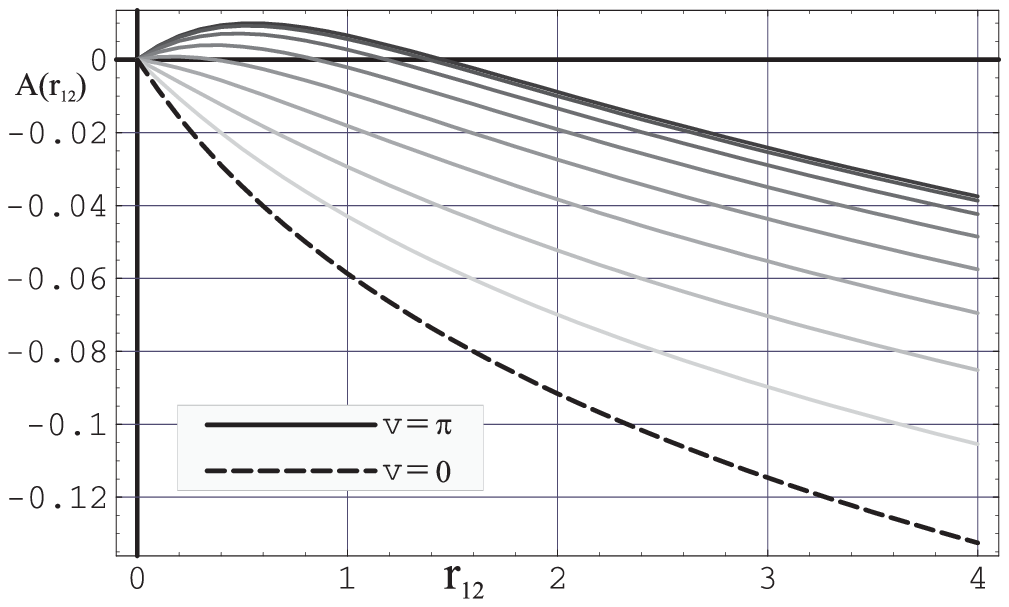}
\epsfxsize=0.5\textwidth \epsfbox{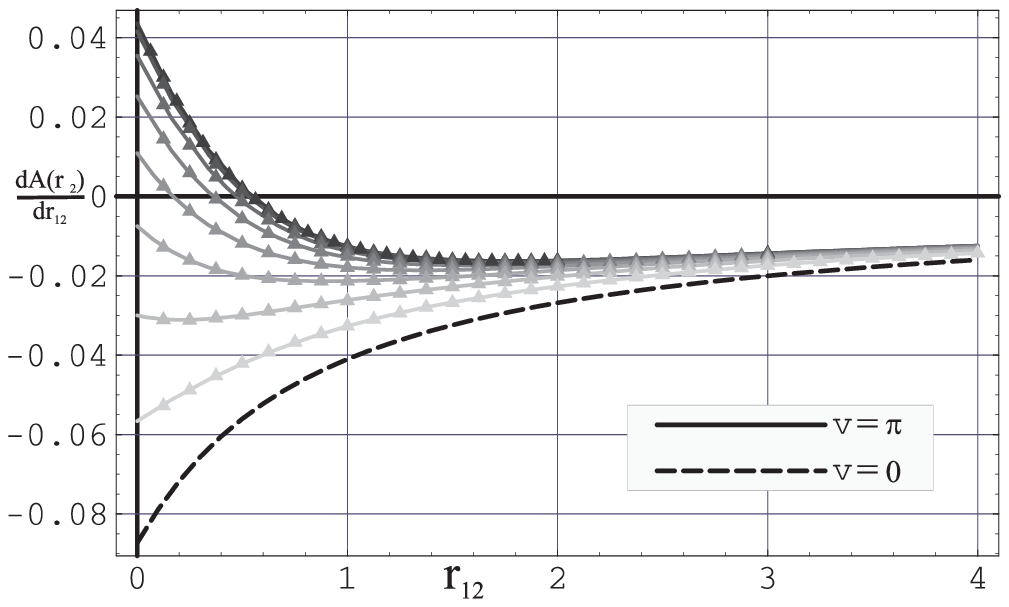}}
\caption{Result of the
numerical evaluation of $A(\v,\D)$. The dashed line corresponds to
the trivial holonomy (left). Fist, we have computed $\d_{\D}
A(\v,\D)$ (right), and then integrated it from $\D=\infty$ to
$\D=0$. All the curves on the left plot go to zero as $\D\rightarrow
0$ with an accuracy $10^{-4}$; that proves the consistency of our
small $\D$ and large $\D$ asymptotics. }
\end{figure}

One can hardly expect a compact analytical expression for the
determinant at arbitrary separations (or instanton sizes) and
holonomies as even for the case of the trivial holonomy, studied by
Gross, Pisarski and Yaffe~\cite{GPY}, it is not known. In this
section we present our results of the numerical evaluation of the
determinant for arbitrary separations and holonomies.

We start from \eq{dvDet} and calculate the exact analytical
expression for the vacuum current (\ref{defJ}) using (\ref{green12})
and the exact expression from \cite{DGPS} for the ADHM construction.
The resulting expression is too lengthy for printing but can be
provided by request. We calculate the trace in the integrand in the
r.h.s. of \eq{dvDet} with a particular choice of
the parameter $\cal P$.

We have taken ${\cal P}=\D$. Using the axial symmetry we can reduce
the number of integrations to 3. We have managed to perform the
integration numerically with the precision $10^{-4}$, despite the
complexity of the expression. The numerical data are shown in
Fig.~3. One can see that it is consistent with all our analytical
results. Following \cite{GPY} we denote
\beq
\log\Det(-\nabla^2)=\left.\log\Det(-\nabla^2)\right|_{\v,T=0}
+A(\v,\D)+\left[P\left(\frac{\bv}{2}\right)-\frac{\pi^2}{12}\right]V+P''\left(\frac{\bv}{2}\right)\frac{\pi\D}{2}\,,
\eeq
where $\left. \log\Det(-\nabla^2)\right|_{\v,T=0}$ is given in \eq{Dinst}.
Note that if \eq{regcontr1} is exact, only the singular part of
the current $J^{\rm s}$ contributes to $A(\v,\D)$, thus $A(\v,\D)$
must be symmetric under the exchange $\v\leftrightarrow\bv$. We have
verified this symmetry numerically. From
(\ref{r12corr},\ref{singcontr}) we see that $A(\v,\D)$ has the
following asymptotics (valid for all $\v$)
\beq
A(\v,\D)=\frac{\log(2\pi)}{6}-\frac{\v\log\v}{12\pi}-\frac{\bv\log\bv}{12\pi}+\frac{1}{18}
-\frac{\gamma}{6}-\frac{\pi^2}{216}-\frac{\log
(\D/\pi)}{12}+\O\left(\frac{1}{\D}\right)
=\frac{(3\v\bv-2\pi^2)\D}{72\pi}+\O\left(\D^{3/2}\right)
\la{Aasy}\eeq
We find that $A(\v,\D)$ can be fitted by the following
expression
\beq
A(\v,\D)\simeq-\frac{1}{12}\log\left(1+\frac{\pi\D}{3}\right)
-\frac{\D\alpha}{216\pi(1+\D)} +\frac{0.00302\;
\D(\alpha+9 \v\bv)}{2.0488+\D^2}
\label{fit}\eeq
where $\alpha=18\v\log\v+18\bv\log\bv-216.611$. This expression has a maximum
absolute error $5\times 10^{-3}$.

\section{Summary
% and discussion
}
In this paper we have considered finite temperature QCD with light fundamental fermions and
calculated the fermionic contribution to the 1-loop quantum weight of the instanton with
non-trivial value of holonomy (or Polyakov line) at spatial infinity.

Finite temperature theory is formulated in terms of the Euclidean QCD partition function
with (anti-)periodic boundary conditions in time with the period $\beta=1/T$, where $T$ is
the temperature.

The instanton with non-trivial holonomy (or the KvBLL caloron~\cite{KvB,LL} for brevity)
is the most general self-dual solution of the YM equations of motion with a unit topological
charge and the above-mentioned periodicity conditions imposed. The solution is parameterized
by the holonomy $\v$ at spatial infinity, the temperature $T=1/\beta$, the size $\D$,
center position, and color orientation. When $\D \ll \beta$ the solution reduces to the ordinary
BPST instanton with size $\rho=\sqrt{\D T/\pi}$ in the domain $x_\mu^2\sim \rho^2$ which
gives the main contribution to the action density. In the opposite limit $\D \gg \beta$
it can be described as a superposition of two (for $N_c=2$) properly gauge-combed
BPS monopoles with the separation $\D$ between them.

One can assume that the effective gauge coupling is reasonably small at the deconfinement and chiral
symmetry restoration temperatures and that therefore the semiclassical ensemble of the KvBLL calorons
makes sense and can be relevant for describing those phase transitions. It is therefore important to calculate
the quantum weight of an individual KvBLL caloron as the first step in the study of the ensemble of calorons.
The gluonic contribution to the quantum weight has been calculated in Ref.~\cite{DGPS}. In this paper we have
extended the methods and the result of that work to the theory with light fermions -- by computing the
fermionic determinant over non-zero modes $\Det' (i \nablaslash)$. The zero fermion modes
\cite{Cherndb} should be taken separately when constructing the ensemble,
in the spirit of Ref.~\cite{DP2a,DP2b}; it will be considered elsewhere.

The main result of the paper is the determinant in the fundamental representation in the background
field of the KvBLL caloron (\ref{APvBP}):
\beq
\half \log \Det'(i \nablaslash(A^P)) = \left.\log\Det(-\nabla^2)\right|_{\v,T=0}+A(\v/T,\D T)
+\left[P\left(\frac{\bv}{2 }\right)-\frac{\pi^2}{12}T^3\right]V
+ P''\left(\frac{\bv}{2}\right)\frac{\pi\D}{2} \,,
\la{resu1}\eeq
where $V$ is a $3d$ volume and $\bv=2 \pi T - \v$.

Since the fundamental representation is sensitive to the sign of the asymptotic holonomy $e^{i \v T \tau_3/2}$,
the fermionic determinant is $4 \pi T$ periodic in $\v$, in contrast to the period of $2 \pi T$ for the gluon
and ghost determinants. We have also considered the background field (\ref{APvBA}) for which one has to exchange
$\v$ with $\bv$ in \eq{resu1}.

The numerical values of the quantity $A(\v,\D)$ are tabulated in Appendix B; they can be fitted by \eq{fit}
consistent with the asymptotic expansion of this quantity at large dyons' separations,
\beqa
\la{r12corrA}
A(\v,\D)&=&\frac{\log (2\pi)}{6}-\frac{\v\log\v}{12\pi}-\frac{\bv\log\bv}{12\pi}+\frac{1}{18}
-\frac{\gamma}{6}-\frac{\pi^2}{216}-\frac{\log (\D/\pi)}{12} \nn \\
&&- \frac{1}{12\D\pi}\left(\log(\v\bv\D^2/\pi^2)-\frac{23\pi^2}{72}+2\gamma+\frac{37}{6}\right)
-\frac{1}{24\D^2\pi}\left(\frac{1}{\v}+\frac{1}{\bv}\right)\\
\nn&+&\frac{1}{72\D^3\pi }\left(\frac{1}{\v^2}+\frac{1}{\bv^2}\right)
-\frac{60-\pi^4}{8640\D^4\pi}\left(\frac{1}{\v^3}+\frac{1}{\bv^3}\right)
+\frac{12-\pi^4}{2880\D^5\pi}\left(\frac{1}{\v^4}+\frac{1}{\bv^4}\right)
+\O\left(\frac{1}{\D^6}\right)\;,
\eeqa
and with the expansion at small $\D$ (see \eq{small_rho}),
\beq
A(\v,\D)=\frac{\D \left(3 \v\bv-2\pi^2\right) }{72\pi } + {\cal O}(\D^{3/2})
\eeq

The generalization of these results to the case of larger groups, $N_c>2$, will be published elsewhere~\cite{GSN}.

%   It modifies the quantum weight of caloron with non-trivial holonomy (found in \cite{DGPS} for the case without fermions)
%by the factor $\left (\Det(-\nabla^2) \right)^2$  and, in particular, it modifies the background
%perturbative potential to $P(\v)-2 N_f P(\half \v + \pi)$  and linear quantum interaction of dyons, which becomes
%\beq
%U_{int}=N_f \pi \D P''\left[\half \v+\pi\right] - 2 \pi \D P''\left[\v\right]
%\eeq

\section*{Acknowledgements}

We thank Dmitry Diakonov and Victor Petrov for useful and
inspirating discussions, and for careful and critical reading
of the manuscript.
Both authors also thank the foundation of non-commercial programs
`Dynasty' for partial support. This work was partially supported by
RSGSS-1124.2003.2.

\appendix

\section{Derivation of the vacuum current in the BPS dyon background}
Here we calculate the isospin-1/2 current (\ref{defJ}) in the M-dyon background (\ref{Mdyon}).
In Section \ref{nearcores} it has been explained that the calculation in the L-dyon background
is equivalent to the calculation for the M-dyon, but with the periodical boundary conditions.
To calculate the vacuum current, we use the expression for the ADHM quantity $|v\>$ for the M-dyon
from the Appendix A of \cite{DGPS}. It states
\beq
|v\>=e^{i\frac{\phi}{2}\tau^3}e^{-i\frac{\pi-\theta}{2}\tau_2}
e^{i\frac{\phi}{2}\tau^3} \sqrt{\frac{\v s}{\sinh (\v s)}}\exp[z\v(i x_4+s\tau^3)]
\la{ADHMM}\eeq
We shall use the explicit expression for the Green function in the ADHM background (\ref{green12}).

\subsection{Singular part of the dyon current $J^\s_\mu$}
The regularized singular part of the current was in fact already computed in Appendix C of \cite{DGPS}. It was
named $j_\mu$ there. The result was
\beqa
\nonumber J^\s_r &= &0,\\
\nonumber J^\s_\phi &= & -\frac{i\v\left(s^2\v^2{\mathrm{csch}^2(s\v )
+ s\v\coth (s\v)-2}\right)}{48\pi^2 s^2\sinh(s\v)}(\cos(\phi) \, \tau_1+\sin(\phi) \,\tau_2),\\
\nonumber J^\s_\theta &= & -\frac{i\v\left(s^2\v^2{\mathrm{csch}^2(s\v )
+ s\v\coth (s\v)-2}\right)}{48\pi^2 s^2\sinh(s\v)}(\sin(\phi) \, \tau_1-\cos(\phi)\, \tau_2 ),\\
J^\s_4 &= &-\frac{i\left( \sinh^2(s\v)-s^3\v^3\coth(s\v)\right)}{48\pi^2 s^3 \sinh^2(s\v)}\tau_3 \;,
\la{Jsmon}\eeqa
where $\tau$ are Pauli matrices, $s$, $\phi$ and $\theta$ are the spherical coordinates centered
at the monopole position. \Eq{Jsmon} obviously does not depend on the temperature and the type of the
boundary conditions as it is a purely zero-temperature object.

\subsection{Regular part of the monopole current $J^\r_\mu$}

We are going to calculate the part of the anti-periodical vacuum current corresponding to
\beq
({\cal G}^{\r})^{ab}(x,y)\equiv\sum_{n\neq 0}
(-1)^n\frac{\<v(x)|v(y_n)\>}{4\pi^2(x-y_n)^2},
\qquad y_n=\vec y,\quad y_{n4}=y_4+n,
\label{Gr}\eeq
This derivation is similar to that done for isospin-1 in \cite{DGPS}.
We repeat it here because it is more simple for the isospin-1/2 case owing to the simple form
of the isospin-1/2 Green function in the general ADHM background.

We define
\beq
\nonumber
J^\r_\mu = J^{\r1}_\mu+J^{\r2}_\mu,\qquad
J^{\r1}_\mu = A_\mu {\cal G}^{\r} + {\cal G}^{\r}A_\mu,\qquad
J^{\r2}_\mu = (\d^x_\mu-\d^y_\mu){\cal G}^{\r} \;.
\eeq

Let us first consider $J^{\r1}_\mu$. We have to compute ${\cal G}^{\r}$ with equal arguments.
Substituting (\ref{ADHMM}) into (\ref{Gr}) one has
\beq
{\cal G}^{\r}(x,x)\equiv\sum_{n\neq 0}(-1)^n\int dz \frac{s\v[\cosh(2 s \v z)1_2
+\sinh(2 s \v z)\tau_3]}{4\pi^2n^2\sinh(s\v)}e^{i n \v z}\;.
\label{Grxx}\eeq
To compute the sum in this expression we use the summation formula (note that $\v<2\pi$):
\beq
\sum_{n\neq 0}\frac{e^{i z n}}{4\pi^2 n^2}
= \frac{z^2}{8\pi^2}-\frac{|z|}{4\pi}+\frac{1}{12},\qquad -2\pi<z<2\pi \;,
\eeq
in particular
\beq
\sum_{n\neq 0}(-1)^n\frac{e^{i z n\beta}}{4\pi^2
n^2\beta^2}=  \frac{z^2}{8\pi^2}-\frac{1}{24\beta^2},\qquad -\pi<z\beta<\pi \;.
\la{Asum}\eeq
It remains now to integrate over $z$. The result is
\beqa
\nonumber
{{\cal G}^\r}(x,x)&=&\frac{(3\v^2-4\pi^2)s^2-6\v \coth(s\v)s+6}{96\pi^2 s^2}\;.
\eeqa

We now turn to the $J^{\r2}_\mu$ part of the current, where
we have to sum over $n$ the derivative of the propagator. First of
all we consider derivatives of the trace in (\ref{Gr}). One finds
for $x= y$:
\begin{widetext}
\beqa \nn(\d^x_\theta-\d^y_\theta)\<v(z)|v(z)\> &\!\!\!= &\!\!\!
ie^{i n \v z}\frac{s\v}{\sinh(s\v)}(\tau_1\sin\phi-\tau_2\cos\phi)\,,
\\
(\nonumber\d^x_\phi-\d^y_\phi)\<v(z)|v(z)\> &\!\!\!= &\!\!\!
i e^{i n \v z}\frac{s\v}{\sinh(s\v)}\cos\theta(\tau_1\cos\phi+\tau_2\sin\phi)\,,
\\
\nn(\d^x_4-\d^y_4)\<v(z)|v(z)\> &\!\!\!= &\!\!\!
-2i e^{i n \v z}\frac{s\v^2 z}{\sinh(s\v)}[1_2\cosh(2 s \v z)+\tau_3\sinh(2 s \v z)]\,,
\\
(\d^x_s-\d^y_s)\<v(z)|v(z)\> &\!\!\!= &\!\!\!0 \;.
\eeqa
\end{widetext}

The derivative of the denominator of (\ref{Gr}) is zero for
$x=y$ except for the derivative with respect to $x_4$, but in this
case we have the expression of the form of \eq{Grxx} with
$n^3/4$ instead of $n^2$ in the denominator. Now we can sum over
$n$. We use the summation formula
\beq
\sum_{n\neq 0}\frac{e^{izn}}{i\pi^2 n^3}=\frac{z^3}{6\pi^2}
-\frac{z|z|}{2\pi}+\frac{z}{3},\;\;\;\;\;-2\pi<z<2\pi  \;.
\eeq
%or equivalently
%\beq
%\sum_{n\neq 0}(-1)^n\frac{e^{izn}}{i\pi^2 n^3}=\frac{z^3}{6\pi^2}
%-\frac{z}{6},\;\;\;\;\;-\pi<z<\pi  \;.
%\eeq
Next one has to integrate over $z$. Combining all pieces we obtain:
\beqa
\la{monopoleRegACurrent}
J^{\rm r}_r&= &0,\\
\nonumber J^{\rm r}_\phi&= &-\frac{i \v(3-3 s\v\coth(s\v)
+s^2 \v^2)}{48\pi^2 s^2\sinh(s\v)}(\sin\phi \,\tau_2+\cos\phi\,\tau_1),\\
\nonumber J^{\rm r}_\theta&= &-\frac{i \v(3-3 s\v\coth(s\v)
+s^2 \v^2)}{48\pi^2 s^2\sinh(s\v)}(\sin\phi \,\tau_1-\cos\phi\,\tau_2),\\
\nonumber J^{\rm r}_4&= &\frac{\v\coth(s\v)(3-3 s\v\coth(s\v)+s^2\v^2)}{48\pi^2 s^2}i\tau_3
+\frac{1-s\v\coth(s\v)}{12 s}i\tau_3.
\eeqa
We have used spherical coordinates. For example, the projection of
$\vec J$ onto the direction
$\vec{n}_\theta= (\cos\theta\cos\phi,\cos\theta\sin\phi,-\sin\theta)$
is denoted by $J_\theta$.

The calculations of the regular part of the periodical vacuum current
is very similar to that of the antiperiodical one. We give here only the result:
\beqa
\la{monopoleRegPCurrent}
\nonumber J^{{\rm r}+}_r&=&0,\\
\nonumber J^{{\rm r}+}_\phi&=&J^{{\rm r}}_\phi+i\v\frac{s\v
-2\tanh\left(\frac{s\v}{2}\right)}{16\pi \sinh(s\v) s}(\sin\phi \,\tau_2+\cos\phi\,\tau_1),\\
\nonumber J^{{\rm r}+}_\theta&=&J^{{\rm r}}_\theta+i\v\frac{s\v
-2\tanh\left(\frac{s\v}{2}\right)}{16\pi \sinh(s\v) s}(\sin\phi\,\tau_1-\cos\phi\,\tau_2),\\
\nonumber J^{{\rm r}+}_4&= &J^{{\rm r}}_4-\frac{\left(1-s\v\coth\left(\frac{s\v}{2}\right)\right)
(1-s\v\coth(r\v))}{8\pi s^2}i\tau_3+\frac{1-s\v\coth(s\v)}{4 s}i\tau_3.
\eeqa

\subsection{Total monopole current $J_\mu$}

Summing up the contributions \eq{Jsmon} and \eq{monopoleRegACurrent} we get for the anti-periodical
boundary conditions
\beqa
\la{monopoleCurrent}
\nonumber J_r&= &0,\\
\nonumber J_\phi&= &-\frac{i \v(1-s\v\coth(s\v))^2}{48\pi^2 s^2\sinh(s\v)}(\sin(\phi) \,\tau_2+\cos(\phi)\,\tau_1),\\
\nonumber J_\theta&= &-\frac{i \v(1-s\v\coth(s\v))^2}{48\pi^2 s^2\sinh(s\v)}(\sin(\phi) \,\tau_1-\cos(\phi)\,\tau_2),\\
\nonumber J_4&= &\frac{1-s\v\coth(s\v)}{12s}i\tau_3-\frac{(1-s\v\coth(s\v))^3}{48\pi^2s^3}i\tau_3,
\eeqa

\newpage

\section{Results of the numerical evaluation}

In this appendix we give the numerical data used to draw Fig.3.
\begin{center}
\begin{tabular}{|l|l|l|}
 \hline
$\v$ & $\D$ & $\d A/\d\D$\\ \hline\hline
% $\pi$ & 3.& -0.01432\\
$\pi$ & 2.& -0.01617\\
   $\pi$  & 1.937& -0.01625 \\   $\pi$ &
   1.875& -0.01629 \\   $\pi$ & 1.8125& -0.01632 \\   $\pi$ & 1.75& -0.01633 \\   $\pi$ &
    1.6875& -0.01631 \\   $\pi$ & 1.625& -0.01625 \\   $\pi$ &
   1.5625& -0.01616 \\   $\pi$ & 1.5& -0.01605 \\   $\pi$ & 1.4375& -0.01588 \\   $\pi$ &
    1.375& -0.01567 \\   $\pi$ & 1.3125& -0.01538 \\   $\pi$ & 1.25& -0.015 \\   $\pi$ &
   1.1875& -0.01455 \\   $\pi$ & 1.125& -0.01399 \\   $\pi$ &
   1.0625& -0.01332 \\   $\pi$ & 1.& -0.01251 \\   $\pi$ & 0.9375& -0.01155 \\   $\pi$ &
   0.875& -0.01017 \\   $\pi$ & 0.8125& -0.00903 \\   $\pi$ & 0.75& -0.00744 \\   $\pi$ &
    0.6875& -0.00556 \\   $\pi$ & 0.625& -0.00338 \\   $\pi$ &
   0.5625& -0.00086 \\   $\pi$ & 0.5& 0.00206 \\   $\pi$ & 0.4375& 0.00542 \\   $\pi$ &
   0.375& 0.00924 \\   $\pi$ & 0.3125& 0.01357 \\   $\pi$ &  0.25 & 0.01847 \\   $\pi$ &
   0.1875& 0.02398 \\   $\pi$ & 0.125& 0.03002 \\   $\pi$ & 0.0625& 0.03655 \\
\hline
  7/8 $\pi$ & 2.& -0.01626\\  7/8 $\pi$ & 1.875& -0.01645\\  7/8 $\pi$ &
    1.75& -0.01643\\  7/8 $\pi$ & 1.625& -0.01637\\  7/8 $\pi$ &
    1.5& -0.01622\\  7/8 $\pi$ & 1.375& -0.01553\\  7/8 $\pi$ &
    1.25& -0.01519\\  7/8 $\pi$ & 1.125& -0.01423\\  7/8 $\pi$ &
    1.& -0.01283\\  7/8 $\pi$ & 0.875& -0.01078\\  7/8 $\pi$ &
    0.75& -0.00793\\  7/8 $\pi$ & 0.625& -0.00404\\  7/8 $\pi$ & 0.5&
    0.00122\\  7/8 $\pi$ & 0.375& 0.00813\\  7/8 $\pi$ & 0.25& 0.0171\\  7/8 $\pi$ &
    0.125& 0.02828\\
\hline
\end{tabular}
\begin{tabular}{|l|l|l|}
 \hline
$\v$ & $\D$ & $\d A/\d\D$\\ \hline\hline
  3/4 $\pi$ & 2.& -0.01652\\  3/4 $\pi$ & 1.875& -0.01664\\  3/4 $\pi$ &
    1.75& -0.01665\\  3/4 $\pi$ & 1.625& -0.01673\\  3/4 $\pi$ &
    1.5& -0.01661\\  3/4 $\pi$ & 1.375& -0.01633\\  3/4 $\pi$ &
    1.25& -0.01581\\  3/4 $\pi$ & 1.125& -0.01498\\  3/4 $\pi$ &
    1.& -0.01375\\  3/4 $\pi$ & 0.875& -0.01197\\  3/4 $\pi$ &
    0.75& -0.0095\\  3/4 $\pi$ & 0.625& -0.00596\\  3/4 $\pi$ &
    0.5& -0.00134\\  3/4 $\pi$ & 0.375& 0.00486\\  3/4 $\pi$ & 0.25&
    0.01291\\  3/4 $\pi$ & 0.125& 0.02311\\ \hline

  5/8 $\pi$ & 3.& -0.01469\\  5/8 $\pi$ & 2.875& -0.01496\\  5/8 $\pi$ &
    2.75& -0.01527\\  5/8 $\pi$ & 2.625& -0.01556\\  5/8 $\pi$ &
    2.5& -0.01585\\  5/8 $\pi$ & 2.375& -0.01613\\  5/8 $\pi$ &
    2.25& -0.01637\\  5/8 $\pi$ & 2.125& -0.01668\\  5/8 $\pi$ &
    2.& -0.01692\\  5/8 $\pi$ & 1.875& -0.01713\\  5/8 $\pi$ &
    1.75& -0.01729\\  5/8 $\pi$ & 1.625& -0.01737\\  5/8 $\pi$ &
    1.5& -0.01736\\  5/8 $\pi$ & 1.375& -0.01721\\  5/8 $\pi$ &
    1.25& -0.01687\\  5/8 $\pi$ & 1.125& -0.01631\\
 5/8 $\pi$ &
    1.& -0.01508\\  5/8 $\pi$ & 0.875& -0.01402\\  5/8 $\pi$ &
    0.75& -0.01208\\  5/8 $\pi$ & 0.625& -0.00935\\  5/8 $\pi$ &
    0.5& -0.00565\\  5/8 $\pi$ & 0.375& -0.00065\\  5/8 $\pi$ & 0.25&
     0.00594\\  5/8 $\pi$ & 0.125 & 0.01445\\ \hline
      1/2 $\pi$ & 2.& -0.0176\\  1/2 $\pi$ & 1.875& -0.01789\\  1/2 $\pi$ &
    1.75& -0.01813\\  1/2 $\pi$ & 1.625& -0.01838\\  1/2 $\pi$ & 1.5& -0.01852\\  1/2 $\pi$ &
     1.375& -0.01859\\  1/2 $\pi$ & 1.25& -0.01854\\
     1/2 $\pi$ &
    1.125& -0.01832\\
\hline
\end{tabular}
\begin{tabular}{|l|l|l|}
 \hline
$\v$ & $\D$ & $\d A/\d\D$\\ \hline\hline
          1/2 $\pi$ & 1.& -0.01793\\  1/2 $\pi$ & 0.875& -0.0171\\  1/2 $\pi$ &
    0.75& -0.01593\\  1/2 $\pi$ & 0.625& -0.01422\\  1/2 $\pi$ & 0.5& -0.01178\\  1/2 $\pi$ &
     0.375& -0.00846\\  1/2 $\pi$ & 0.25& -0.00377\\  1/2 $\pi$ & 0.125 & 0.00255\\ \hline
      3/8 $\pi$ & 3.& -0.01545\\  3/8 $\pi$ & 2.875& -0.01581\\  3/8 $\pi$ &
    2.75& -0.01618\\  3/8 $\pi$ & 2.625& -0.01657\\  3/8 $\pi$ &
    2.5& -0.0169\\  3/8 $\pi$ & 2.375& -0.01728\\  3/8 $\pi$ &
    2.25& -0.01777\\  3/8 $\pi$ & 2.125& -0.0182\\  3/8 $\pi$ &
    2.& -0.01854\\  3/8 $\pi$ & 1.875& -0.01905\\  3/8 $\pi$ &
    1.75& -0.01947\\  3/8 $\pi$ & 1.625& -0.01989\\  3/8 $\pi$ &
    1.5& -0.02027\\  3/8 $\pi$ & 1.375& -0.02065\\  3/8 $\pi$ &
    1.25& -0.02099\\  3/8 $\pi$ & 1.125& -0.02122\\  3/8 $\pi$ &
    1.& -0.02131\\  3/8 $\pi$ & 0.875& -0.02131\\  3/8 $\pi$ &
    0.75& -0.02119\\  3/8 $\pi$ & 0.625& -0.0207\\  3/8 $\pi$ &
    0.5& -0.01985\\  3/8 $\pi$ & 0.375& -0.01842\\  3/8 $\pi$ &
    0.25& -0.01609\\  3/8 $\pi$ & 0.125 & -0.0128\\
\hline
      1/4 $\pi$& 3.& -0.01616\\  1/4 $\pi$& 2.875& -0.01658\\  1/4 $\pi$&
    2.75& -0.01702\\  1/4 $\pi$& 2.625& -0.01747\\  1/4 $\pi$&
    2.5& -0.01795\\  1/4 $\pi$& 2.375& -0.01855\\  1/4 $\pi$&
    2.25& -0.01909\\  1/4 $\pi$& 2.125& -0.01956\\  1/4 $\pi$&
    2.& -0.02014\\  1/4 $\pi$& 1.875& -0.02079\\  1/4 $\pi$&
    1.75& -0.02148\\  1/4 $\pi$& 1.625& -0.022\\  1/4 $\pi$&
    1.5& -0.02278\\  1/4 $\pi$& 1.375& -0.02349\\  1/4 $\pi$&
    1.25& -0.02439\\  1/4 $\pi$& 1.125& -0.02525\\
\hline
\end{tabular}
\begin{tabular}{|l|l|l|}
 \hline
$\v$ & $\D$ & $\d A/\d\D$\\ \hline\hline
 1/4 $\pi$&
    1.& -0.02629\\  1/4 $\pi$& 0.875& -0.02707\\  1/4 $\pi$&
    0.75& -0.02806\\  1/4 $\pi$& 0.625& -0.02896\\  1/4 $\pi$&
    0.5& -0.02986\\  1/4 $\pi$& 0.375& -0.03064\\  1/4 $\pi$&
    0.25& -0.03121\\  1/4 $\pi$& 0.125& -0.03107\\ \hline
      1/8 $\pi$& 4.& -0.01428\\  1/8 $\pi$& 3.875& -0.01462\\  1/8 $\pi$&
    3.75& -0.01487\\  1/8 $\pi$& 3.625& -0.01523\\  1/8 $\pi$&
    3.5& -0.0156\\  1/8 $\pi$& 3.375& -0.01602\\  1/8 $\pi$&
    3.25& -0.01645\\  1/8 $\pi$& 3.125& -0.01693\\  1/8 $\pi$&
    3.& -0.0174\\  1/8 $\pi$& 2.875& -0.01797\\  1/8 $\pi$&
    2.75& -0.01851\\  1/8 $\pi$& 2.625& -0.01908\\  1/8 $\pi$&
    2.5& -0.01972\\  1/8 $\pi$& 2.375& -0.02033\\  1/8 $\pi$&
    2.25& -0.02105\\  1/8 $\pi$& 2.125& -0.02182\\  1/8 $\pi$&
    2.& -0.02266\\  1/8 $\pi$& 1.875& -0.02353\\  1/8 $\pi$&
    1.75& -0.02448\\  1/8 $\pi$& 1.625& -0.02557\\  1/8 $\pi$&
    1.5& -0.02672\\  1/8 $\pi$& 1.375& -0.0279\\  1/8 $\pi$&
    1.25& -0.02939\\  1/8 $\pi$& 1.125& -0.03094\\  1/8 $\pi$&
    1.& -0.03265\\  1/8 $\pi$& 0.875& -0.03462\\  1/8 $\pi$&
    0.75& -0.03677\\  1/8 $\pi$& 0.625& -0.03923\\  1/8 $\pi$&
    0.5& -0.04202\\  1/8 $\pi$& 0.375& -0.04517\\  1/8 $\pi$&
    0.25& -0.04876\\  1/8 $\pi$& 0.125& -0.05272\\\hline
& & \\
& & \\
& & \\
& & \\
& & \\
& & \\
& & \\
& & \\
\hline
\end{tabular}
\\ The absolute error in these data is less than $10^{-4}$.
\end{center}

\end{document}